\documentclass[aps,twocolumn,showpacs,superscriptaddress]{revtex4}

\usepackage{graphicx}
\usepackage{amsfonts}
\usepackage[figuresright]{rotating}  
\usepackage{amssymb}
\usepackage{amsmath}
\usepackage{psfrag}
\usepackage{subfigure}
\usepackage{multirow}
\usepackage{tabularx}

\def\avg#1{\langle#1\rangle}

\def\nn{\nonumber}

\def\beq{\begin{eqnarray}}
\def\eeq{\end{eqnarray}}

\begin{document}

\title{Spontaneous inhomogeneous phases in ultracold dipolar Fermi gases}

\author{Kai Sun}
\affiliation{Condensed Matter Theory Center and Joint Quantum Institute,
Department of Physics, University of Maryland, College Park, MD 20742, USA}
 
\author{Congjun Wu}
\affiliation{Department of Physics, University of California, 
San Diego, CA 92093, USA}

\author{S. Das Sarma}
\affiliation{Condensed Matter Theory Center and Joint Quantum Institute, 
Department of Physics, University of Maryland, College Park, MD 20742, USA}

\begin{abstract}
We study the collapse of ultracold fermionic gases into inhomogeneous 
states due to strong dipolar interaction in both 2D and 3D. 
Depending on the dimensionality, 
we find that two different types of inhomogeneous states are
stabilized once the dipole moment reaches a critical value $d>d_c$:
the {\it stripe phase} and {\it phase separation} between high and low 
densities. 
In 2D, we prove that the stripe phase is always favored for $d\gtrsim d_c$, 
regardless of the microscopic details of the system. 
In 3D, the one-loop perturbative calculation suggests that the same 
type of instability leads to phase separation. 
Experimental detection and finite-temperature effects are discussed.
\end{abstract}
\date{\today}
\maketitle 
\section{Introduction}
The rapid experimental development of ultracold atomic and molecular physics
has opened up new opportunities to study quantum many-body systems with
electric and magnetic dipolar interactions \cite{ospelkaus2008,ni2008,
griesmaier2005,chicireanu2006,mcclelland2006,lu2010,koch2008,
lahaye2009,menotti2007}.
An important feature of dipolar interaction is its explicit
spatial anisotropy of the $d_{r^2-3z^2}$-type when dipole moments
are aligned by external fields.
For the fermionic dipolar systems, $^{40}$K-$^{87}$Rb has been cooled 
down almost to quantum-degeneracy \cite{ospelkaus2008}.
Anisotropic Fermi liquid theories of the single
particle and collective properties have been investigated 
\cite{sogo2008,miyakawa2008,ronen2009,chan2010,fregoso2009,lin2010,Kestner2010}.
Furthermore, unconventional Cooper pairing structures have
been studied, including the $p_z$-channel pairing in the
single component systems \cite{baranov2002,baranov2004,baranov2008a,
you1999,bruun2008,zhao2009}, and the competition between the $s+d$-wave 
singlet and the $p_z$-wave triplet channels \cite{samokhin2006,wu2010,shi2009}.
In particular, a novel pairing state of the $s+ip$ type broken
time-reversal symmetry has been pointed out \cite{wu2010}. 
Moreover, magnetic dipolar systems of fermions have also 
been experimentally realized \cite{lu2010}.
Exotic states of the ferro-nematic Fermi liquid and unconventional
magnetic states have been predicted \cite{fregoso2010,fregoso2009a}.

Fermionic systems can spontaneously break translational
symmetry in both charge and spin channels.
Many years ago, Overhauser pointed out that even in the weak coupling
regime of the interacting electron gas, a $2k_f$ spin-density wave state
always wins over the uniform paramagnetic state at the Hartree-Fock
level \cite{Overhauser1962}.
However, correlation effects may suppress this instability and such a
state has not been experimentally confirmed.
Recently, quantum liquid crystal phases in strongly correlated systems
have been intensively studied, particularly in doped Mott
insulators \cite{kivelson1998}. Stripe ordering has been observed 
in high T$_c$ cuprates, other transition metal oxides, and quantum 
Hall systems at high Landau levels \cite{kivelson1998,kivelson2003,
Tranquada2005,fradkin2009}.

In this article, we study the instability toward the spontaneous 
inhomogeneous phase in dipolar fermionic systems.
In two dimensional (2D) systems, the strongest density-channel
instability occurs at non-zero momentum, which drives the density 
wave states under strong dipolar interactions.
This effect is based on the peculiar feature of the Fourier transform of 
the dipolar interaction, which is robust against microscopic details.
However in three dimensions (3D), the instability at the one-loop level
occurs at zero momentum, thus it leads to phase separation
into high and low density regions.

This paper is organized as follows: In Sec. \ref{sec:model}, we construct
the model Hamiltonian for a single-component dipolar Fermi gas and
present the calculation of the static susceptibility of density fluctuations via 
diagrammatic expansion. Then, we study the instability of the homogeneous
ground state in 2D and 3D in Secs.~\ref{sec:2D} and \ref{sec:3D}
correspondingly. Here, we present both a nonperturbative conclusion, which is 
based on the analytic property of the susceptibility and the leading-order 
perturbative calculation, which is use to testify the nonperturbative  conclusion. 
The experimental detection is then discussed in Sec. \ref{sec:exp}. 
Finally, we conclude our paper by studying the contributions of thermal fluctuations 
at finite temperature and the effects of higher-order terms in Sec. \ref{sec:discussion}.

\section{Model}\label{sec:model}
We consider the single-component dipolar Fermi gas with dipole
moment aligned by an external electric field.
The long-distance physics of this system is described by the following Hamiltonian
\begin{align}
H=\sum_{\vec{k}} (\epsilon_{\vec k}-\mu) c_{\vec k}^\dagger c_{\vec k}
+\sum_{\vec k,\vec k^\prime,\vec q}V(\vec q)c_{\vec k+\vec q}^\dagger 
c_{\vec k} c_{\vec k^\prime-\vec q}^\dagger c_{\vec k^\prime},
\label{eq:model}
\end{align}
with $\epsilon_{\vec{k}}$ being the energy dispersion relation and $\mu$ 
the chemical potential. 
In 2D, the dipolar interaction in the momentum space $V_{\textrm{2D}}(\vec q)$ takes the form \cite{chan2010},
\begin{align}
V_{\textrm{2D}}(\vec q)=2\pi d^2 
P_2(\cos\theta)[\frac{1}{\epsilon}-q]+\pi q d^2 \sin^2\!\theta  
\cos 2 \phi_q,
\label{eq:interaction}
\end{align}
where $d$ is the dipole moment; $\phi_q$ is the azimuthal angle of the momentum $\vec q$ and 
$\theta$ is the angle between the direction of the dipoles and the $z$-axis.
$\epsilon$ is a short-range cutoff roughly equal to the thickness of the system along the $z$-direction.
Equation \eqref{eq:interaction} is valid for $q\epsilon\ll 1$. However, as shown below, the main conclusion of this paper only relies on the value of  $V(\vec{q})$ at small $q$, and hence all the
qualitative results remain invariant even if more accurate and more complicated interactions were
taken into account. We shall also emphasize here that for $\theta=0$, $V_{\textrm{2D}}(\vec q)$ is
invariant under $\textrm{SO}(2)$ rotations. However, for any $\theta\ne 0$, the rotational symmetry 
is reduced to two-fold ($C_2$). This symmetry property is crucial as a system undergoes the
transition to a charge-density-wave (CDW) phase. As will be discussed below,
in a system with $\textrm{SO}(2)$ symmetry, the CDW instability usually leads to the formation
of a triangular lattice~\cite{Brazovskii75} (a 2D Wigner crystal). On the other hand, the same instability  
in a system with $C_2$ symmetry is compatible with the formation of a  unidirectional
CDW, i.e., a stripe phase~\cite{Sun08}.

Notice that Eq. \eqref{eq:interaction} contains an isotropic part which 
depends on the microscopic cut-off $\epsilon$ and an anisotropic part which
depends on the azimuthal angle of $\phi_q$.
We first consider the purely anisotropic case at
\begin{align}
\theta=
\theta_0=\cos^{-1}\frac{1}{\sqrt{3}},
\end{align}
and then the interaction simplifies
into
\begin{align}
V_{\textrm{2D}}(\vec q)=\frac{2\pi d^2}{3}  q \cos 2 \phi_q.
\label{eq:int}
\end{align}
The major feature of Eq. \eqref{eq:int} is its linear dependence
on $q$, which play important roles in driving the instability at 
non-zero wavevectors. 

\begin{figure}
\begin{center}
\includegraphics[width=0.45\textwidth]{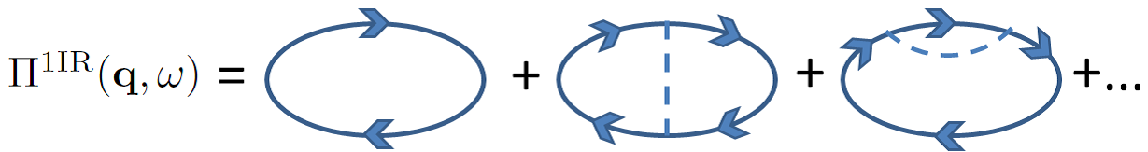}
\end{center}
\caption{(online color) Diagrammatic expansion for $\Pi_{\textrm{2D}}^\textrm{1IR}(\vec q)$. 
The solid lines are free-fermion propagators
and the dashed lines represent the dipolar interaction.}
\label{fig:diagram}
\end{figure}

The stability of a system against density fluctuations is determined 
by the static susceptibility of density fluctuations:
\begin{align}
\Pi_{\textrm{2D}}(\vec q,\omega=0)=\avg{\rho(\vec q,\omega=0)
\rho(-\vec q,\omega=0)},
\label{eq:pi}
\end{align}
where 
\begin{align}
\rho(\vec q,\omega)=\sum_{\vec k,\Omega} c^\dagger_{\vec k+\vec q,
\Omega+\omega}c_{\vec k,\Omega},
\end{align}
is the density operator.
The inverse of $\Pi_{\textrm{2D}}(\vec q,\omega=0)$ is the 
energy gap $\Delta_{\textrm{2D}}(\vec q)$ 
of creating a density wave at momentum $\vec q$.
Using the technique of the diagrammatic expansion, 
$\Delta_{\textrm{2D}}(\vec q)$ can be computed order by order as
\begin{align}
\Delta_{\textrm{2D}}(\vec q)&=\Pi_{\textrm{2D}}(\vec q,\omega=0)^{-1}\nn\\
&=[\Pi_{\textrm{2D}}^\textrm{1IR}(\vec q,\omega=0)]^{-1}+V_{\textrm{2D}}(\vec q),
\label{eq:gap}
\end{align}
where
$\Pi^\textrm{1IR}(\vec q,\omega)$ is the one-particle-irreducible 
density-density correlation \cite{Amit1984,Mahan1990} shown in 
Fig. \ref{fig:diagram}.
The stability criteria of the homogeneous phase corresponds
to the energy gap $\Delta_{\textrm{2D}}(\vec q)$ being positive-definite
for any $\vec q$.
Otherwise, the creation of a density wave could lower the total energy of the system leading to 
the condensation of density waves.
Obviously, $\Delta_{\textrm{2D}}(\vec q)$ is positive-define for any $\vec q$
in the noninteracting Fermi gas.
This stability remains for a weakly-interacting Fermi liquid. 
On the other hand, if the interaction has an attractive channel 
[e.g. the dipolar interaction in Eq. \eqref{eq:int}], an inhomogeneous state 
could take advantage of the attractive interaction.
Under strong enough attractive interactions, i.e., $V_{\textrm{2D}}\rightarrow -\infty$,
the homogeneous ground state is unstable.

\section{stripe phases in 2D}\label{sec:2D}

We begin from the weak coupling side, by tuning up the interaction strength.
In general, there is a critical dipole moment $d_c$ 
such that $\Delta_{\textrm{2D}}(\vec q)>0$ at any $\vec q$ for $d<d_c$, 
but $\Delta_{\textrm{2D}}(\vec q)<0$ at some $\vec q$ for $d>d_c$.
At $d=d_c$, $\Delta_{\textrm{2D}}(\pm \vec Q)=0$ for the momenta 
$\pm\vec Q$ and remain positive at other momenta \cite{note:degeneracy}. 
If we increase $d$ further above $d_c$, the density wave fluctuations 
with momentum $\vec q\sim \vec Q$ undergo an instability
and condense. Depending on whether $\vec Q$ is $0$ or not, 
this instability has two different fates. 
If $Q>0$, the condensation carries a nonzero momentum, which 
leads to a stripe state: a unidirectional charge-density-wave state with wavevector $\vec Q$.
On the other hand, if $\vec Q=0$, the condensation takes 
place at an infinite wavelength and results in phase separation
between high and low density regions. Here, the 
typical size of these high (low) density regions and their spatial 
arrangements are determined by the microscopic 
details of the system. In the particular case of a dipolar gas, 
the phase separation is sensitive to the details of the short-range behaviors of the interaction.

\subsection{$\theta=\theta_0$}

\begin{figure}
\begin{center}
\vspace{-4mm}
\includegraphics[trim = 0mm 0mm 0mm 20mm,width=0.4\textwidth]{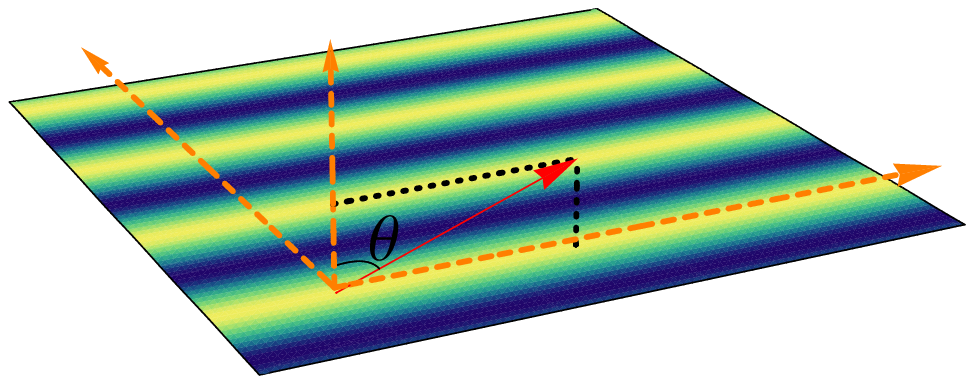}
\end{center}
\caption{(online color) The stripe phase in 2D dipolar gases. The three dotted arrows 
mark the $x$, $y$ and $z$ axes.
The lighter (darker) regions in the 2D plane represent higher (lower) 
density. The solid line with an arrow 
represents the direction of the external field, which aligns the 
direction of the dipoles.}
\label{fig:stripe}
\end{figure}

Due to the 2D space-inversion symmetry $\vec r\rightarrow -\vec r$, 
$[\Pi_{\textrm{2D}}^\textrm{1IR}(\vec q,\omega)]^{-1}$ 
is an even function of $\vec q$. As a result, the  Taylor expansion of 
$\Delta_{\textrm{2D}}(\vec q)$ at small momentum
must take the following form
\begin{align}
	\Delta_{\textrm{2D}}(\vec q)=&[\Pi_{\textrm{2D}}^\textrm{1IR}
(\vec q=0)]^{-1}+ \frac{2\pi d^2}{3} q \cos 2 \phi_q
	\nonumber
	\\&+\frac{1}{2} \sum_{i,j} \partial_{q_i}\partial_{q_j}
[\Pi_{\textrm{2D}}^\textrm{1IR}(\vec q)]^{-1}|_{\vec q =0} q_i q_j
	+\ldots
\label{eq:expansion} 
\end{align}
%We focus on the most unstable direction $\phi_q=\frac{\pi}{2}$, in which 
%the attraction in $V_\textrm{2D}(\vec q)$ is the strongest and
%the linear $q$ term in Eq. \eqref{eq:expansion} has a negative coefficient.
At small momentum, $q\sim 0$, Eq.~\eqref{eq:expansion} was dominated by 
the first two terms. And hence the point $\vec q=0$ is a local saddle point
of $\Delta_{\textrm{2D}}(\vec q)$ (i.e., a local minimum along the 
line of $\phi_q=0$ and a local maximum along $\phi_q=\pi/2$). Since
$q=0$ is a saddle point, instead of the minimum point of $\Delta_{\textrm{2D}}(\vec q)$, 
as $d$ increases, $\Delta_{\textrm{2D}}(\vec q)$ at some finite 
momentum will become negative before $\Delta_{\textrm{2D}}(\vec q=0)$ 
could reach zero. Thus, $Q$ must be a nonzero value,
which indicates that the system will collapse into a stripe phase 
for $d>d_c$, instead of phase separating. This is the main conclusion of
this paper. It is noteworthy that this conclusion only depends on the
small-momentum (i.e. long-range) behaviors of the interaction $V(\vec{q})$.
Therefore, the stripe instability is a universal 
property of the 2D dipolar Fermi gas at $d>d_c$, insensitive to the
microscopic details of the system. In 2D electron gases, a similar
conclusion has been found in the classical limit at high temperature 
when the quantum fluctuations are negligible \cite{spivak2004}.

We emphasize that %this conclusion is independent of microscopic details.
%More importantly, 
the conclusion of $Q>0$ is a non-perturbative result 
which remains valid to any order in the loop expansion of 
Fig. \ref{fig:diagram}. The only assumption required here is the 
analyticity of $\Pi_{\textrm{2D}}^\textrm{1IR}(\vec q,\omega=0)$ 
as a function of $\vec q$ near $\vec q=0$. 
In a Fermi liquid, it is well-known that $\Pi^\textrm{1IR}(\vec q,\omega=0)$ 
is non-analytic at $\vec q=2 k_F$ where $k_F$ is the Fermi wavevector. 
However, there is no reason of a non-analytic behavior at small momentum 
in any Fermi liquid to our knowledge.
The higher-order diagrams in Fig. \ref{fig:diagram} contain interaction 
lines which are non-analytic at small momenta.
However, due to the fact that dipolar interactions are 
short-ranged in 2D, we do not expect any singularity
for $\Pi_{\textrm{2D}}^\textrm{1IR}(\vec q,\omega=0)$ around 
$\vec q=0$.

\begin{figure}
\begin{center}
\subfigure[]{\includegraphics[width=0.3\textwidth]{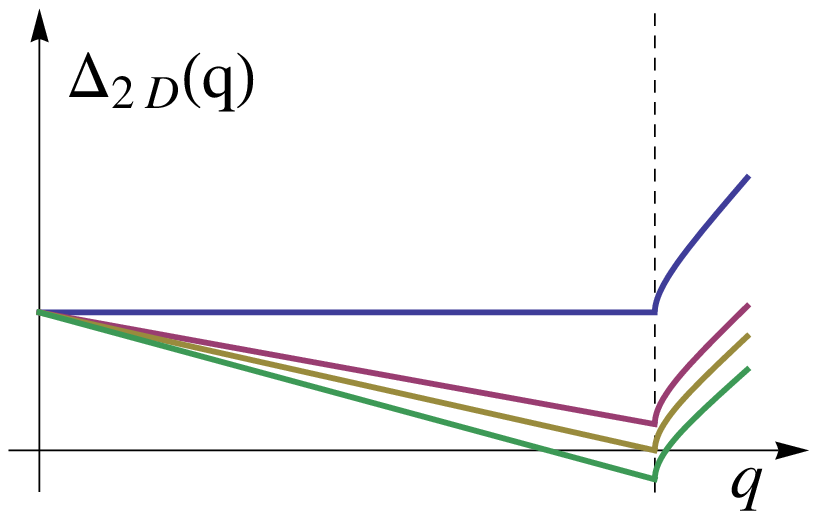}
\label{fig:delta2D}}
\subfigure[]{\includegraphics[width=0.3\textwidth]{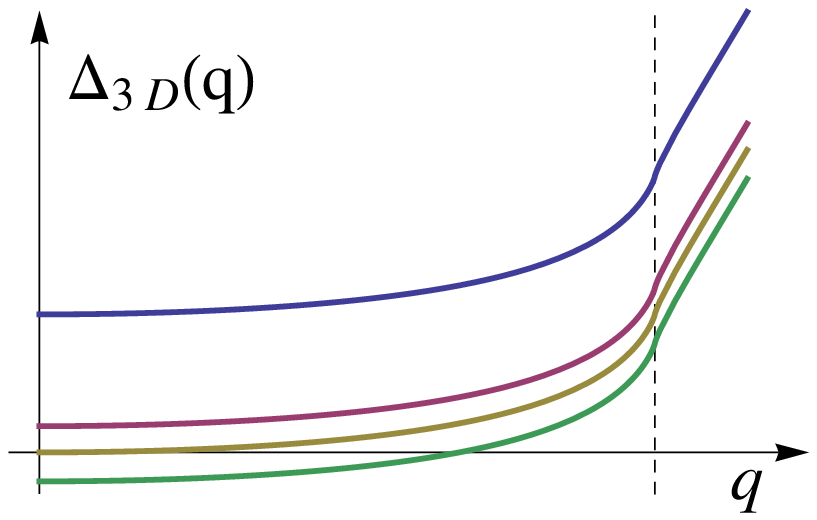}
\label{fig:delta3D}}
\end{center}
\caption{(online color) One-loop calculation of $\Delta(\vec q)$ in (a) $2D$ and (b) $3D$. 
In the $2D$ case, we choose $\phi_{\vec q}=\frac{\pi}{2}$ and for the $3D$, 
$\theta_{\vec q}=\pi/2$.
From top to bottom, the dipole momentum for each of the curves 
are: $d/d_c=0$, $0.9$, $1$, and $1.1$ respectively. 
The dashed vertical line marks the momentum $2 k_F$.
}
\label{fig:Delta}
\end{figure} 

We check this conclusion using the loop expansion in Fig. \ref{fig:diagram}
to the one-loop level for a %continuum 
2D system at $T=0$ in continuum
($\epsilon_{\vec k}=k^2/2 m$ with $m$ being the mass of the particles)
and find 
\begin{align}
\Delta_{\textrm{2D}}(\vec q)=\frac{2\pi}{m}
[\frac{1}{1-\sqrt{1-\frac{1}{x^2}}\eta(1-x)}+
\frac{d^2}{d_c^2} x \cos 2\phi_{\vec q}], 
\end{align}
where $x=q/2 k_F$ and $d_c$ is 
the critical dipole momentum $d_c=\sqrt{3/(2 k_F m)}$. 
$\phi_{\vec q}$ is the azimuthal angle of the momentum $\vec q$ and 
$\eta(x)$ is a step function with $\eta(x)=1$ for $x>0$ and
$\eta(x)=0$ for $x<0$. As shown in Fig. \ref{fig:delta2D}, the order wavevector
$Q=2 k_F$. In experiments, it is often more convenient to vary the density for 
a constant $d$. There, the density need to be higher than a critical value
to reach the stripe phase. For KRb ($d=0.57$ D)
the stripe phase requires the interparticle spacing to be smaller than $\sim 10^{-4}$ cm, 
and for LiCs ($d\sim5.5$ D), the critical spacing is $\sim  10^{-2}$ cm. 
We emphasize that $Q$ and $d_c$ we found here are based on the leading order
approximation, and can be renormalized by higher order corrections.
However, the fact that $Q>0$ remains valid, which is one of the key conclusion of this work.

\subsection{$\theta \ne \theta_0$}

For $\theta \ne \theta_0$, the same conclusion applies for the single
component case.
The $\epsilon$-term in Eq. \eqref{eq:interaction} is momentum independent,
which corresponds to short-range contact interaction.
Thus it has no effect on the single component fermions, and hence
\begin{align}
V(q)=q\pi d^2[-2P_2(\cos\theta)+\sin^2\theta\cos2\phi_q].
\end{align}
Along $\phi_q=\frac{\pi}{2}$ which remains the direction of the strongest
instability, 
\begin{align}
V(q)=-2 q\pi d^2 \cos^2 \theta,
\end{align}
which is negative and grows linearly
with $q$ for an arbitrary value of $\theta$. Thus, the
above conclusion still holds.
We notice that even for $\theta=0$, i.e., when the dipoles are perpendicular
to the plane, the stripe instability still exists at large values of $d$
although the interaction is now repulsive. For such an isotropic system (at $\theta=0$),
it is often assumed that three charge density waves form a triangular lattice to 
minimize the breaking of the rotational symmetry \cite{Brazovskii75}.
However, if the rotational symmetry is spontaneously broken before the homogeneous
state becomes unstable, a stripe phase can be stabilized at zero temperature \cite{Sun08}.
In this case, the stripe ordering spontaneously 
breaks both the rotational symmetry and the translational symmetry
in the direction perpendicular to the stripes, while only the
latter is spontaneously breaking in the stripe phases at
$\theta\ne 0$. Due to its unique symmetry breaking pattern, the 
low-energy properties of the stripe phase at $\theta=0$ is described by 
the quantum McMillan-de Gennes theory, which is fundamentally different from other
similar phases at $\theta\ne 0$ \cite{Sun08}.
As a result, the stripe phase at $\theta=0$ is unstable
at any finite temperature due to thermal fluctuations \cite{Sun08,Chaikin98}, 
while the stripe phases at $\theta\ne0$ has a power-law quasi-long range order blow
a transition temperature $T_{KT}$ (See below).

\section{fermions in 3D and multicomponent fermions}\label{sec:3D}
In this section, we study two more classes of systems with similar dipolar interactions:
(a) dipolar gases in 3D and (b) 2D dipolar gases composed by multicomponent fermions.
 
In 3D, the dipolar interaction is
\begin{align}
V_\textrm{3D}(\vec q)=\frac{8\pi d^2}{3} P_2(\cos \theta_{\vec q}),
\label{eq:V3D}
\end{align}
with $\theta_{q}$ measuring the angle between $\vec q$ and the dipole moment.
Due to the structure of $V_\textrm{3D}$, the linear term of $\vec q$ in 
Eq. \eqref{eq:expansion} 
is now absent in 3D, and hence we could not exclude the possibility 
of $\vec Q$ being zero. 
In fact, the one-loop calculation at $T=0$ for a system in continuum shows that
\begin{align}
\Delta_{\textrm{3D}}(\vec q)=
\frac{4\pi^2}{m k_F}[&(1+\frac{1-x^2}{2 x}
\log |\frac{1+x}{1-x}|)^{-1}	
\nn\\&
+\frac{d^2}{d_C^2}(3 \cos \theta_{\vec q}^2-1)],
\label{eq:interaction3D}
\end{align}
where $x=q/2 k_F$ and $d_c=\sqrt{6\pi/(k_F m)}$.
As can be seen from Fig. \ref{fig:delta3D}, we found $\vec Q=0$
within the one-loop approximation,  implying a phenomenon of phase separation 
for $d>d_c$.  As discussed above, this is sensitive to the short-range details 
of the interaction and no universal conclusion is available for case.

Now we briefly discuss the multi-component case in 2D (e.g. the degrees of freedom of 
hyperfine spins) whose Hamiltonian is
\begin{align}
H=&\sum_{\vec{k},\sigma} (\epsilon_{\vec k}-\mu) c_{\vec k,\sigma}^\dagger c_{\vec k,\sigma}
\nn\\
&+\sum_{\vec k, \vec k^\prime,\vec q}\sum_{\sigma,\sigma'}V(\vec q)c_{\vec k+\vec q,\sigma}^\dagger 
c_{\vec k,\sigma} c_{\vec k^\prime-\vec q,\sigma'}^\dagger c_{\vec k^\prime,\sigma'},
\label{eq:model2}
\end{align}
where the spin indices $\sigma$ and $\sigma'$ run from $-N/2$ to $+N/2$ for
some integer $N$.
Here, $V(\vec q)$ remains the same as the single-component case in Eqs. \eqref{eq:interaction} and
\eqref{eq:V3D}.
For a single-component dipolar gas, due to Pauli exclusion principle, the $\epsilon$-term 
in Eq. \eqref{eq:interaction} (short-range interactions) has no contribution to any physical 
properties of the system. However, for multi-component Fermi gases, this term naturally exists in the 
inter-component interactions and suppresses the stripe instability if it is repulsive.
For simplicity we only consider the case of $N=2$. 
For $\theta>\theta_0$ ($\theta<\theta_0$), this short-range part is attractive (repulsive), and hence the stripe state become unfavorable for $\theta<\theta_0$.
Furthermore, such an finite momentum instability does not exist in 
the spin channel because the spin channel interaction arises 
from the exchange interaction which is regular for small vector
of $\vec q$.

\section{Experimental Detection}
\label{sec:exp}
The stripe phase has a density modulation in the direction perpendicular 
to the stripes, which can be detected directly via the measurement of 
the local densities.  
In addition, any scattering experiments would show two interference Bragg 
peaks at momentum $\pm \vec Q$ in a stripe phase due to this density 
modulation.

There are also other indirect tools to detect a stripe phase.
Considering a density wave with density varying along the $y$ axis , whose density profile 
(to the leading order) is 
\begin{align}
\rho(x,y)=\rho_0+\rho_1 \cos(Q y+\phi),
\end{align}
with $(x,y)$ being the 2D coordinate in the real space, if we place
this system in a shallow potential trap and then introduce an extra narrow 
1D potential well along the direction of the stripe, for example,
\begin{align}
V(x,y)=-\alpha \delta(y-y_0),
\label{eq:pinning}
\end{align}
it is energetically favorable to have $y=y_0$ being a high density region due to the potential well $V(x,y)$. We emphasize that in reality, this 1D potential well of $V(x,y)$ shall have a finite width in
the $y$ direction. However, as long as the width is much smaller than the wavelength of
the stripe, one can treat it as a $\delta$-function to the leading order approximation.
If we move the potential trap along the $x$ direction but keep the 
location of the potential well $V(x,y)$ unchanged, the stripes will 
largely remain at their initial positions if the displacement of the trap 
is less than half the wave-length of the stripes $\pi/\vec Q$, and the 
stripes will jump over a distance of $2\pi/\vec Q$ if the displacement 
becomes larger than $\pi/\vec Q$ to same the potential energy.
On the other hand, the jump would not happen in a homogeneous state. 
This jump will lead to a center of mass oscillation for the trapped
particles due to the lack of dissipation, which could serve as an 
indirect signature for the stripe phase.

This technique is in close analogy to the  ``pinning'' effect, which was used to 
experimentally distinguish the nematic phase 
from a putative stripe phase in two-dimensional electron gases (2DEG) under 
high magnetic field~\cite{Cooper2001,Cooper2002,Cooper2003}. 
There, the impurities in the crystal provide similar 
effects as the local potential in Eq. \eqref{eq:pinning}, which will pin down 
the stripe if the stripe exists. In principle, the presence of stripes will leads
to the pinning effect, which shall result in nonlinear signal in the $I$-$V$ curve at 
low bias. However, in the 2DEG under high magnetic field, the absence of such 
a nonlinear $I$-$V$ curve was interpreted as the absence of the stripe ordering.
Different from the 2DEG, in the case we studied here, the similar pinning effect
would lead to oscillations for a system with stripe ordering, due to the absence of 
dissipation in ultracold dipolar gases.

\section{Discussions}
\label{sec:discussion}
Since a real system has a finite temperature, the contribution of the 
thermal fluctuations comes into play.
Their effects can be studied based on the symmetry-breaking pattern of 
the ordered phase. For a 2D system in the continuum, the homogeneous 
phase at $d<d_C$ has continuous translational and two-fold rotational 
symmetries (rotation by $\pi$) at $\theta\ne 0$. 
For $d>d_C$, the stripe phase spontaneously breaks the continuous
translational symmetry in one direction and leads to one 
gapless Goldstone mode as required by symmetry. 
In a 2D system with $d>d_C$, if we increase the temperature from $T=0$, 
such a Goldstone mode will destroy the long-range order of the stripe phase. 
However, a quasi-long-range power-law correlation would still remain for 
temperature below a transition temperature $T_\textrm{KT}$ \cite{Chaikin98}.
For $T<T_\textrm{KT}$, although the real-space density oscillation of
the stripes is hard to observe due to the lack of a true long-range order, 
the density profile in Fourier space will have two peaks located at 
$\pm \vec Q$ which decays as a power-law function of $|\vec q\mp\vec Q|$. 
Above $T_\textrm{KT}$, the correlation becomes short-ranged and the phase 
transition at $T=T_\textrm{KT}$ belongs to the usual Kosterlitz$-$Thouless (KT) 
university class. 

For dipolar molecules  in a 2D optical lattice, if the wavevector of the 
stripe phase is commensurate with the lattice wavevector, the stripe phase 
breaks no continuous symmetry and hence no gapless Goldstone mode is
present. Therefore, the long-range order of the stripe phase remains at
finite $T$ below a critical temperature $T_C$. 
However, if the stripe phase is incommensurate with the underlying 
lattice, a KT transition is expected.

In the study of Secs. \ref{sec:2D} and \ref{sec:3D}, we were focusing on
on the susceptibility shown in Eq. \eqref{eq:pi}. Effectively, this procedure is 
equivalent and closely related to the liner-response approximation. 
Thus, the higher-order terms beyond the liner-response 
approximation were ignored in our study.
These higher order terms are highly nontrivial and could become nonlocal at 
certain wave-vector. For example, the free energy contains terms 
proportional to $\rho(\vec q)^{5/2}$ or $\rho(\vec q)^{7/2}$ at $q\sim 2 k_F$, 
as shown in Refs. \cite{Maslov2006} and  \cite{Sun08}. Although these terms could be studied 
perturbatively within certain approximations, there is no available technique to take 
care of them nonperturbatively in a general way. In general, the contributions of these 
higher order terms are non-universal. In the case of a second-order phase transition, from the 
homogeneous phase to the inhomogeneous phase, these higher order term are 
unimportant at low energy and hence all the conclusions above remains. 
However, if the transition is first order, their contributions could become dominant.
For a system with short-range interactions only, a first order transition is usually believed
to leads to phase separation (as in the liquid/gas-solid transition). However, it was also 
shown that a stripe phase was favored in the presence of long-range interactions like 
Coulomb \cite{Emery1993,Kivelson1993} or dipolar interactions \cite{spivak2004}. 
%This effect, where a long-range Coulomb interaction stabilizes the stripe ordering, 
%was referred to  as the Coulomb frustration.

In summary, we find that the 2D dipolar Fermi gas undergoes a stripe 
ordering at large dipolar interaction strength, i.e., the density wave 
instability at finite momentum, instead of phase separation.
For the single component case, this instability occurs regardless
of the dipole orientation. This arises from the fact that the Fourier transform of
the dipolar interaction increases linearly with wavevector.
For the multicomponent case, due to the short-range Hartree interaction,
the stripe instability only exists at $\theta>\theta_0$.

\section*{Note added}
After the completion of first online version of the manuscript, 
we learned the work by Yamaguchi {\it et al.} \cite{Yamaguchi}, in which the 
charge-density-wave instability of dipolar Fermi gas are studied using mean-field 
approximation.

\begin{acknowledgments}
K. S. and S. D. S. are supported by JQI-NSF-PFC and AFOSR-MURI.
C. W. is supported by NSF under No. DMR-0804775, and Sloan Research
Foundation. 
\end{acknowledgments}

%\bibliography{dipolar}

\begin{thebibliography}{10}
\bibitem{note:degeneracy}
The degeneracy between $\vec{Q}$ and $-\vec{Q}$ is due to the
space-inversion symmetry $\vec{r}\rightarrow -\vec{r}$.
\end{thebibliography}

\end{document}